\documentclass[
	aps,
	prl,
	reprint,
	amsmath,amssymb,
	superscriptaddress,floatfix
]{revtex4-2}
\usepackage{graphicx} %graphics handler
\usepackage{dcolumn} %Align table columns on decimal point
\usepackage{bm}% bold math
\usepackage{amssymb,amsmath}
\usepackage{epstopdf}
\usepackage{braket}
\usepackage{color}
\usepackage{xcolor}
\usepackage{mathrsfs}
\usepackage{lipsum}
\usepackage{array}
\usepackage{afterpage}
\usepackage{multirow}
\usepackage[normalem]{ulem}  % to use \sout{}
\usepackage{ragged2e}

\begin{document}
\title{$g$-factor engineering with InAsSb alloys toward zero band gap limit}

\author{Yuxuan Jiang}
\email{yuxuan.jiang@ahu.edu.cn}
\affiliation{School of Physics and Optoelectronics Engineering, Anhui University, Hefei 230601, China}
\affiliation{Center of Free Electron Laser and High Magnetic Field, Anhui University, Hefei 230601, China}
\author{Maksim Ermolaev}
\affiliation{Department of Electrical and Computer Engineering, Stony Brook University, Stony Brook, New York 11794, USA}
\author{Seongphill Moon}
\affiliation{National High Magnetic Field Laboratory, Tallahassee, Florida 32310, USA}
\affiliation{Department of Physics, Florida State University, Tallahassee, Florida, 32306, USA}
\author{Gela Kipshidze}
\affiliation{Department of Electrical and Computer Engineering, Stony Brook University, Stony Brook, New York 11794, USA}
\author{Gregory Belenky}
\affiliation{Department of Electrical and Computer Engineering, Stony Brook University, Stony Brook, New York 11794, USA}
\author{Stefan Svensson}
\affiliation{U. S. Army Research Directorate, 2800 Powder Mill Rd, Adelphi, MD 20783, USA}
\author{Mykhaylo Ozerov}
\affiliation{National High Magnetic Field Laboratory, Tallahassee, Florida 32310, USA}
\author{Dmitry Smirnov}
\affiliation{National High Magnetic Field Laboratory, Tallahassee, Florida 32310, USA}
\author{Zhigang Jiang}
\email{zhigang.jiang@physics.gatech.edu}
\affiliation{School of Physics, Georgia Institute of Technology, Atlanta, Georgia 30332, USA}
\author{Sergey Suchalkin}
\email{sergey.suchalkin@stonybrook.edu}
\affiliation{Department of Electrical and Computer Engineering, Stony Brook University, Stony Brook, New York 11794, USA}
\date{\today}
\begin{abstract}
Band gap is known as an effective parameter for tuning the Land\a'{e} $g$-factor in semiconductors and can be manipulated in a wide range through the bowing effect in ternary alloys. In this work, using the recently developed virtual substrate technique, high-quality InAsSb alloys throughout the whole Sb composition range are fabricated and a large $g$-factor of $g\approx -90$ at the minimum band gap of $\sim 0.1$ eV, which is almost twice that in bulk InSb is found. Further analysis to the zero gap limit reveals a possible gigantic $g$-factor of $g\approx -200$ with a peculiar relativistic Zeeman effect that disperses as the square root of magnetic field. Such a $g$-factor enhancement toward the narrow gap limit cannot be quantitatively described by the conventional Roth formula, as the orbital interaction effect between the nearly triply degenerated bands becomes the dominant source for the Zeeman splitting. These results may provide new insights into realizing large $g$-factors and spin polarized states in semiconductors and topological materials.
\end{abstract}

\maketitle
Land\a'{e} $g$-factor is a major material parameter describing the response of electron spins to an external magnetic field ($B$). In solid state physics, the long-standing interest in finding large $g$-factor materials originates from the peculiar spin-dependent transport and optical phenomena, which hold great promises for potential applications in spintronics \cite{Zutic2004spintronics,Awschalom2007review}, nonreciprocal spin photonics \cite{Sengupta2020electron}, and quantum information processing \cite{Kosaka2001electron,Lutchyn2018review}.

In III-V semiconductors, the electron $g$-factor is known to observe the renowned Roth formula \cite{Roth1959Theory,Pryor2006PRL}
\begin{align*}
    g=g_e-\frac{2}{3}(\frac{1}{E_g}-\frac{1}{\Delta+E_g})E_P,
\end{align*}
where $g_e\approx 2$, $E_g$, $\Delta$, and $E_P$ are the free electron $g$-factor, the band gap, the spin-orbit coupling, and the Kane energy, respectively. The Roth formula is, in principle, a single-band theory, which explains the $g$-factor as a result of remote band perturbations \cite{dresselhaus2007group}. A recent study further reveals the connection between the $g$-factor and the Berry curvature of the bands due to the mixing of wavefunctions \cite{chang2008berry}. Therefore, it is natural to expect a large $g$-factor in narrow band gap materials. Indeed, among all the binary III-V semiconductors, InSb has the smallest band gap and thus the largest $g$-factor, $g\approx -52$ \cite{Jancu2005PRB,isaacson1968electron}.

\begin{table*}[t]
\caption{Composition and thickness of the core layers in the MBE-grown InAsSb samples of different Sb concentrations. The core layer structure is shown in Fig. 1(a).}
\begin{tabular}{c|cc|cc|cc|cc|cc}
\hline
\multicolumn{1}{c|}{Sb (\%)}  & \multicolumn{2}{c|}{Grading (nm)}  &\multicolumn{2}{c|}{Bottom barrier (nm)} &\multicolumn{2}{c|}{Absorber (nm)} &\multicolumn{2}{c|}{Top barrier (nm)}  &\multicolumn{2}{c}{Cap layer (nm)}\\
\hline
9   & Not & N/A
    & Al$_{80}$Ga$_{20}$As$_{6.2}$Sb$_{93.8}$ &500 
    & InAs$_{91}$Sb$_{9}$ &1000
    & Al$_{80}$Ga$_{20}$As$_{6.2}$Sb$_{93.8}$ &200
    & InAs$_{91}$Sb$_{9}$ &100 \\
22   & Al$_{85}$In$_{15}$Sb & 1600
    & Al$_{95}$In$_{4.5}$Sb &500 
    & InAs$_{78}$Sb$_{22}$ &1000
    & Al$_{95}$In$_{4.5}$Sb &200
    & InAs$_{78}$Sb$_{22}$ & 100 \\
44   & Al$_{60}$In$_{40}$Sb & 3000
    & Al$_{68}$In$_{32}$Sb &500
    & InAs$_{56}$Sb$_{44}$ &1000
    & Al$_{68}$In$_{32}$Sb &200
    & InAs$_{56}$Sb$_{44}$ &100 \\
50   & Al$_{39}$In$_{61}$Sb & 2630
    & Al$_{63}$In$_{37}$Sb &250 
    & InAs$_{50}$Sb$_{50}$ &1500
    & Al$_{63}$In$_{37}$Sb &200
    & InAs$_{50}$Sb$_{50}$ &100 \\
60  & Al$_{40}$In$_{60}$Sb &4000
    & Al$_{48}$In$_{52}$Sb &500 
    & InAs$_{40}$Sb$_{60}$ &1000
    & Al$_{48}$In$_{52}$Sb &200
    & Al$_{40}$In$_{60}$Sb &100 \\
\hline
\end{tabular}
\label{table1}
\end{table*}

To further reduce the band gap, one can resort to ternary semiconductor InAsSb alloys, as the bowing effects can suppress the band gaps below those of their binary constituents \cite{Vurgaftman2001Band}. Recent experiments have firmly established a strong negative bowing of the band gap with a bowing coefficient of $\sim 0.8$ eV \cite{Svensson2012bandgap,Suchalkin2016electronic}, leading to a minimum band gap of 0.1 eV when the Sb composition is close to 63\%. As a result, the theoretical estimation of the electron $g$-factor based on the Roth formula reaches as high as $g=-117$, which is more than twice that in InSb \cite{Mayer2020ACS,Goswami2021NL}. Such a large tunable range of band gaps and $g$-factors has rendered InAsSb alloy a promising platform for spintronics \cite{metti2022spin,Mayer2020ACS,Goswami2021NL}, topological phase engineering \cite{Suchalkin2018engineering,Suchalkin2020APL,Winkler2016topological}, and infrared (IR) optoelectronics \cite{rogalski2020inassb,svensson2017materials,donetsky2019inassb}. 

However, there remain concerns about the high expectation value of $g$-factors in InAsSb alloys towards the zero gap limit. On the one hand, the Roth formula is a single-band theory and fails to predict the correct result as the band gap reduces, and multiband theories such as the $k \cdot p$ model are necessary. Also, the experimental studies of $g$-factors in the narrow or zero band gap region, particularly for Dirac materials, do not exhibit extraordinarily large $g$-factors as expected \cite{jiang2022giant,Jiang2017Landau,Jiang2019valley}. On the other hand, there are technical difficulties in obtaining high-quality InAsSb alloys with the Sb composition close to 50\%. Although in earlier works, InAsSb alloys with different alloy compositions were fabricated, they suffered from a large lattice mismatch between the alloy and substrate, which led to the relaxation of the bulk alloy and formation of numerous threading dislocations deteriorating the electronic properties \cite{Smith1992interband,tersoff1993dislocations}. The increased disorder, particularly in the intermediate composition range, can contribute to the extrinsic composition dependence of the key material parameters that determine the $g$-factor in a bulk material, such as $E_g$, $\Delta$, and $E_P$ \cite{siggia1974k} and makes the experimental characterization of their intrinsic electronic property difficult.

Recent advances in the virtual substrate technique allow for the molecular beam epitaxy (MBE) growth of high-quality unstrained, unrelaxed InAsSb alloys in the whole composition range \cite{Suchalkin2016electronic,Belenky2011properties}, providing a perfect opportunity for experimental studies of the material parameters and $g$-factors in the narrow band gap region. In this work, we present a systematic investigation of the band structure evolution with the composition in InAsSb alloys via a combination of magneto-absorption measurements and $k \cdot p$ calculations. We find that the Kane energy shows very little bowing effect across the entire composition range, but the $g$-factor increases significantly as the band gap reaches the minimum. When $E_g \rightarrow 0$, the Landau levels (LLs) of the triply degenerated bands become fully relativistic (i.e., LL energy $\propto \sqrt{B}$) due to the dominant orbital interaction, and their relative wavefunction mixing determines the spin states and energy spacing of the LLs. For typical III-V (more generally, zinc-blende type) semiconductor, we find that these relativistic LLs are highly spin polarized along with maximized energy spacings, which could lead to a $g$-factor of $g\approx -200$ at 1 T (vs. $g\rightarrow -\infty$ based on the Roth formula), overwhelmingly larger than most of the two-band Dirac materials. Our findings may provide a new perspective for $g$-factor engineering in future devices based on semiconductors and topological materials.

\begin{figure}
\includegraphics[width=0.45\textwidth]{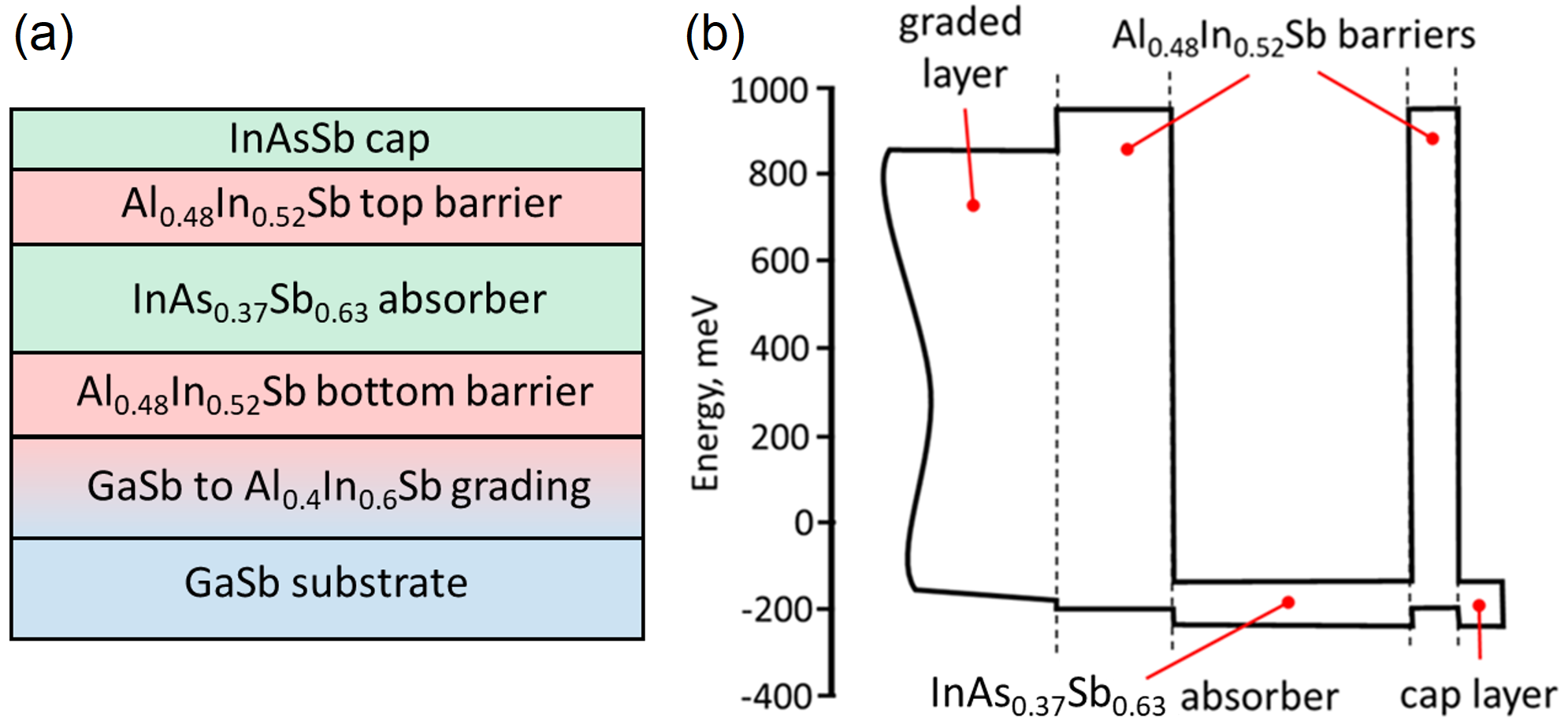}
\caption{(a) Structure layout of the InAs$_{0.37}$Sb$_{0.63}$ sample. The InAsSb alloy (absorber) is sandwiched between the two Al$_{0.48}$In$_{0.52}$Sb barriers. (b) Schematic band alignment of the InAs$_{0.37}$Sb$_{0.63}$ sample as an example. The zero energy corresponds to the top of the GaSb valence band.}
\label{fig1}
\end{figure}

Five InAs$_{1-x}$Sb$_x$ alloy samples are studied in this work, with $x=0.09$, 0.22, 0.44, 0.50, and 0.63. These samples are grown by solid-source MBE on undoped GaSb(100) substrates. The $x=0.50$ sample was grown using VEECO Gen II MBE system in Army Research Laboratory, and the other samples were grown using VEECO GEN930 MBE system in Stony Brook University. The growth process has been described previously in Ref. \cite{Suchalkin2016electronic}. The core structure and band alignment of our InAs$_{0.37}$Sb$_{0.63}$ sample are schematically shown in Fig. \ref{fig1} as an example. Information on the core structures of these samples is summarized in Table \ref{table1}. In addition, samples with $x = 0.09$, 0.22, and 0.44 are n-doped (Te-doped, 2$\times$10$^{16}$ cm$^{-3}$), and samples with $x=0.50$ and 0.63 are grown without intentional doping. To avoid the formation of two-dimensional electron “pockets” due to band bending at the boundaries of the InAsSb layer (absorber), the barriers and cap are p-doped to 10$^{16}$ cm$^{-3}$. The three-dimensional character of the carrier motion in InAsSb is confirmed by magneto-transport measurements in tilted magnetic fields \cite{Suchalkin2016electronic}.

InAsSb alloy samples are then studied with magneto-IR spectroscopy, which is known for its accuracy in determining electronic band structures. The samples are placed inside a superconducting magnet at liquid helium temperature (the effective temperature at the sample is measured to be $T=5$ K). The samples are illuminated with IR radiation in the Faraday configuration using a Bruker 80v Fourier-transform IR spectrometer. A composite Si bolometer is mounted behind the sample to detect the transmitted light signal at different magnetic fields. 

%%%%% FIG. 2 %%%%%
\begin{figure}[t]
\includegraphics[width=8.5cm] {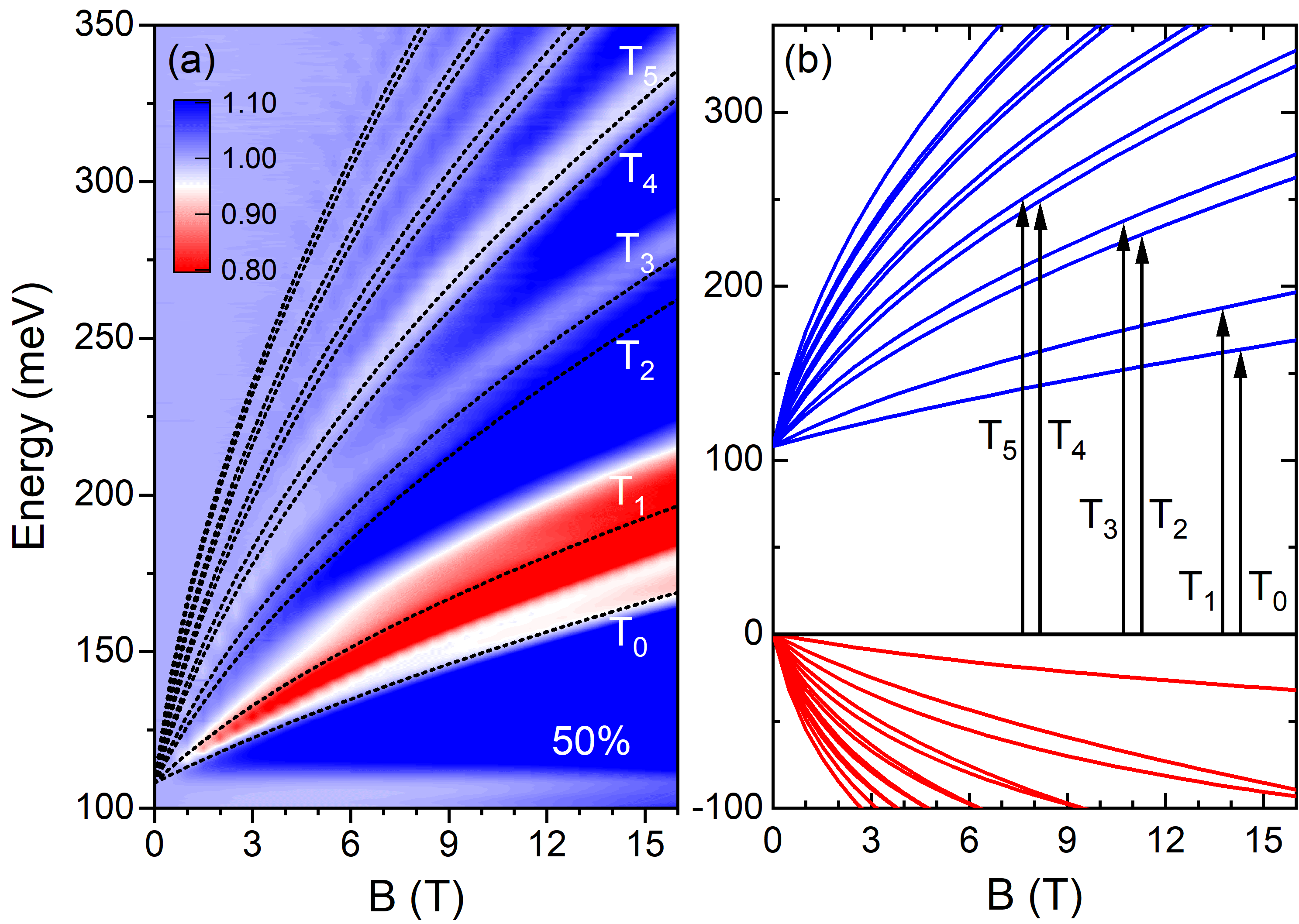}
\caption{(a) False color plot of the normalized transmission $T(B)/T(0\text{T})$ of the InAs$_{0.50}$Sb$_{0.50}$ alloy sample. The dashed lines indicate the fitting results from the $k \cdot p$ model using parameters given in Table \ref{table2}. The first few absorption modes are labeled with $T_i,i=0,1,..., 5$. (b) Calculated LL fan diagram of InAs$_{0.50}$Sb$_{0.50}$ at $\Gamma$ point. The blue, black, and red colors denote the LLs from the EB, HH, and LH bands, respectively. The arrows show the low-lying LL transitions, in correspondence to those in panel (a).}
\label{fig2}
\end{figure}
%%%%%%%%%%%%%%

Figure \ref{fig2}(a) shows the false color plot of the normalized transmission $T(B)/T(0 \text{T})$ of the InAs$_{0.50}$Sb$_{0.50}$ sample as a typical example. A series of absorption modes, which blue-shift in energy with increasing magnetic fields, can be identified and attributed to LL transitions. The low-lying transitions are labeled with $T_0$--$T_5$. These modes originate from the same non-zero energy intercept as the magnetic field approaches zero, indicative of the nature of interband LL transitions. The energy intercept allows for direct readout of the band gap $E_g=108$ meV. 

To quantitatively describe these LL transitions and extract other material parameters, we employ the well-established eight-band $k \cdot p$ model to fit the experimental results \cite{jiang2017probing,jiang2022giant,Sanders2003electronic,Smith1992interband}. The model consists of several parameters, including $E_g$, $\Delta$, $E_P$, the electron effective mass $m^*$, and the modified Luttinger parameters $\gamma_1$, $\gamma_2$, and $\gamma_3$. To simplify the Hamiltonian, we first assume $\gamma_{1,2,3}=0$. Meanwhile, we set $A_c=\hbar^2/2m^*-E_P(3E_g+2\Delta)/6m_0 E_g(E_g+\Delta)=0$, where $\hbar$ is the reduced Planck constant and $m_0$ is the free electron mass, to avoid spurious solutions \cite{Foreman1997elimination}. Finally, we focus on the $\Gamma$ point LLs, which carry the dominant contributions to the observed optical transitions. With these assumptions, the $k \cdot p$ Hamiltonian is greatly simplified while, as we will show below, ensuring a good agreement between the experiment and model. The simplified Hamiltonian now reads
\begin{equation}
H_{k\cdot p}=
\begin{bmatrix}
H_{+} & 0 \\
0 & H_{-}\\
\end{bmatrix},
\end{equation}
where
\begin{align*}
H_{+}=\begin{bmatrix}
E_g & i\sqrt{3}V^{\dagger} & iV&\sqrt{2}V \\
-i\sqrt{3}V & 0 & 0 & 0 \\
-iV^{\dagger} & 0 &0 & 0\\
\sqrt{2}V^{\dagger} &0 & 0 &-\Delta\\
\end{bmatrix},
\end{align*}
\begin{align*}
H_{-}=\begin{bmatrix}
E_g & -\sqrt{3}V & -V^{\dagger} & i\sqrt{2}V^{\dagger} \\
-\sqrt{3}V^{\dagger} & 0 & 0 
 &0\\
-V & 0 &0 &0\\
-i\sqrt{2}V &0 &0  &-\Delta
\end{bmatrix}.
\end{align*}
Here, $V =\frac{1}{\sqrt{6}}P_{0}k_{-}$, $\mathbf{k}=(k_x,k_y,k_z)$ is the wave vector, $k_{\pm}=k_x \pm k_y$, and $P_0$ is related to the Kane energy by $E_P=2 m_0 P_0^2/\hbar^2$. The bases for the Hamiltonian are in the order of the electron band (EB) spin up, heavy hole (HH) spin up, light hole (LH) spin down, split-off (SO) spin down, EB spin down, HH spin down, LH spin up, and SO spin up bands.

To calculate the LL energies, we apply the ladder operator formalism and the following ansatz to the two subblocks of the Hamiltonian \cite{jiang2017probing,Sanders2003electronic}. For $H_+$ subblock, the ansatz is $\ket{n_+}=[\ket{n-1},\ket{n-2},\ket{n},\ket{n}]^{T}$. For $H_-$ subblock, the ansatz is $\ket{n_-}=[\ket{n-1},\ket{n},\ket{n-2},\ket{n-2}]^{T}$. Here, $[...]^T$ denotes the transpose operation, $n$ is a positive integer, and $\ket{n}$ is the $n^{\text{th}}$ harmonic oscillator eigenfunction. Further details of the calculation can be found in Refs. \cite{jiang2017probing,Sanders2003electronic}.

\begin{table}[b]
\begin{center}
\caption{Fitting parameters extracted from experiments using the $k \cdot p$ model.}
\begin{tabular}{cccccc}
    \hline
    Sb & $E_g$(eV) &$\Delta$(eV)  &$E_P$(eV) & $g_{\textrm{exp}}$ &$g_{\textrm{theory}}$\\
    \hline
    0\%  & 0.415 &0.390  &19   &15.0   & 12.8 \\
    9\%  & 0.315 &0.323 &22   &20.0   & 21.6\\
    22\% & 0.220  &0.276 &20   &29.4 & 31.7 \\
    44\% & 0.132 &0.280  &19   &63.2 & 63.2 \\
    50\% & 0.108 &0.300   &19   &76.0   & 87.4\\
    63\% & 0.100   &0.375 &21   &91.5 &108.5  \\
    100\%& 0.235 &0.800   &23.3 &51.3 & 49.1\\
    \hline
  \end{tabular}
  \label{table2}
\end{center}
\end{table}

With the calculated LLs, we can fit the experimental data and extract the corresponding band parameters. The dashed lines in Fig. \ref{fig2}(a) show the best fits to the data, and Fig. \ref{fig2}(b) shows the calculated LL structure using the fitting parameters in Table \ref{table2}. In Fig. \ref{fig2}(b), we also label out the corresponding low-lying LL transitions for $T_0$--$T_5$, where we assume the dominant contributions to the observed transitions in Fig. \ref{fig2}(a) are the HH to EB LL transitions \cite{Smith1992interband}. 

%%%%% FIG. 3 %%%%%
\begin{figure}[t]
\includegraphics[width=8.5cm] {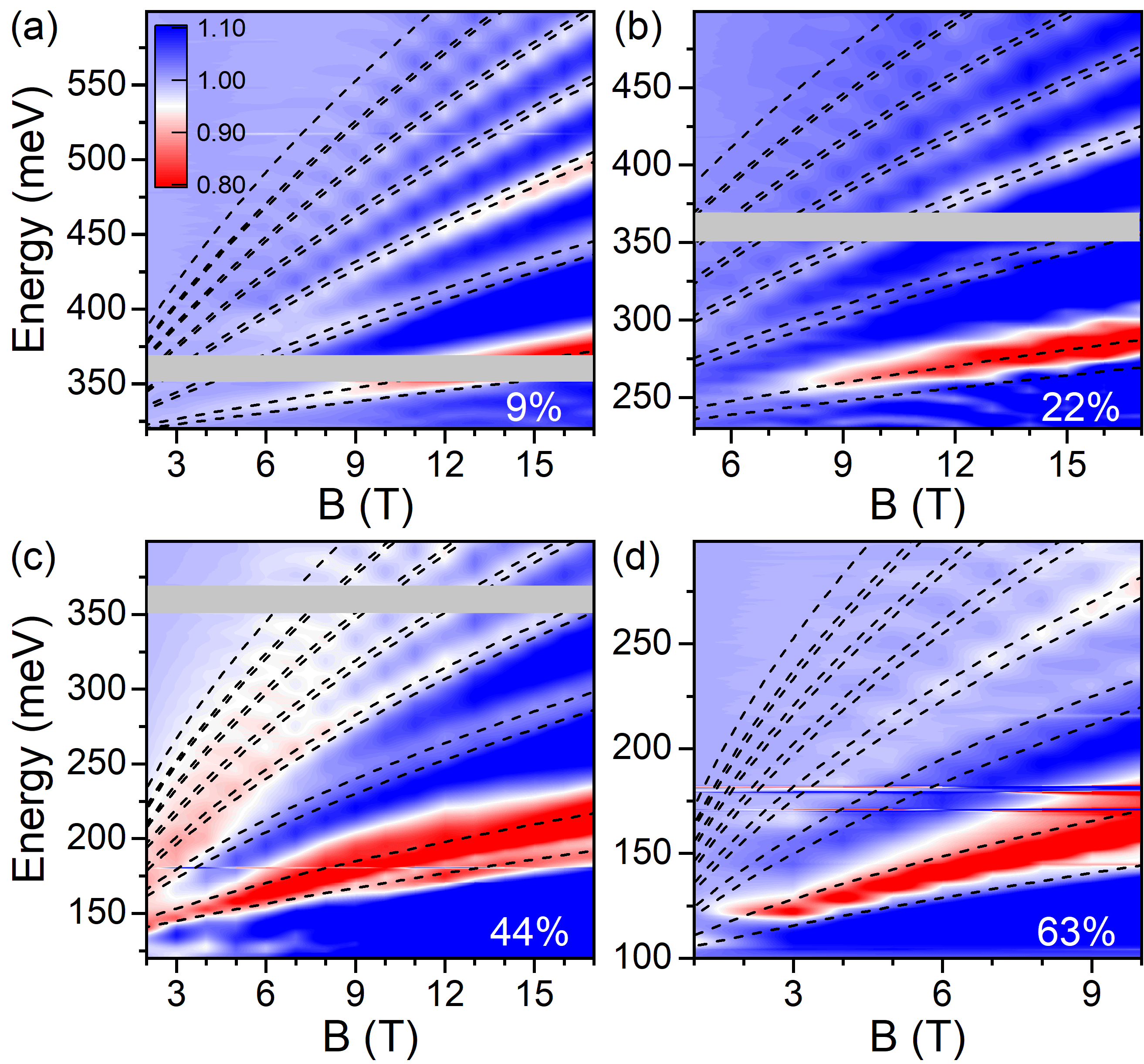}
\caption{(a-d) False color plot of the normalized transmission $T(B)/T(0 \text{T})$ for InAsSb samples of (a) 9\%, (b) 22\%, (c) 44\%, and (d) 63\% Sb compositions. The dashed lines indicate the fitting results from the $k \cdot p$ model using parameters given in Table \ref{table2}. The gray areas are opaque region to IR light and shows no intensity. The color scales in all panels are kept the same.}
\label{fig3}
\end{figure}
%%%%%%%%%%%%%%

Following the above analysis, we can analyze the experimental results of other InAsSb alloys with different Sb compositions. Figure \ref{fig3} shows their false color plot of the normalized transmission data for Sb compositions of $9\%, 
 22\%, 44\%$, and $63\%$, respectively. Similar to Fig. \ref{fig2}(a), the dashed lines are best fits to the data using the $k \cdot p$ model, which exhibits excellent agreement with the experiment. Table \ref{table2} summarizes the band parameters extracted from the fitting for different Sb concentrations. We note that, the actual fitting parameters are $E_g$ and $E_P$, whereas $\Delta$ does not critically affect the fitting results as the SO band is distant from the other bands. Here, we assume that $\Delta$ follows the bowing relation of ternary InAsSb alloys reported in Ref. \cite{Vurgaftman2001Band}.

Based on the results in Table \ref{table2}, we can study the bowing effects of the band parameters. First, the band gap $E_g$ bows positively with the Sb concentration. By comparing the interband LL transition energies of different compositions, we find that the energy decreases as the Sb composition increases and $E_g$ reaches its minimum $\sim 100$ meV at 63\% Sb concentration. The extracted $E_g$ versus Sb composition gives a bowing coefficient of 0.83, consistent with our previous result \cite{Suchalkin2016electronic}. 

Second, the Kane energy $E_P$ shows a weak bowing effect throughout the entire Sb composition range. This is in contrast to an earlier work \cite{Smith1992interband}, where $E_P$ bows significantly with the Sb concentration. It is likely that the samples in Ref. \cite{Smith1992interband} were grown with relaxed strain due to a strong mismatch of the lattice parameters between the substrate and the epilayers, which degraded the quality of the alloys, particularly near the middle of the composition range. According to Ref. \cite{siggia1974k}, this may lead to additional coupling between the conduction and valence bands and hence bowing of $E_P$. 

Lastly, we discuss the bowing effect in $g$-factors. The g-factor for nth LL is defined as $g_n=\min_{m} |E_{n,\uparrow(\downarrow)}(B)-E_{m,\downarrow(\uparrow)}(B)|/B$, where $\min \{...\}$ finds the nearest LL of opposite spin. Based on this definition, the experimental $g$-factors ($g_{\textrm{exp}}$) are extracted from the splitting of the two lowest EB LLs at 1 T, calculated using the $k \cdot p$ model with experimental band parameters. For comparison, we also calculate the theoretical $g$-factors ($g_{\textrm{theory}}$) from the Roth formula. In both cases, we observe a negative bowing. That is, the $g$-factor gradually increases with increasing Sb composition and reaches a maximum when the band gap reaches a minimum at 63\% Sb. Then, the $g$-factor decreases with increasing band gap and Sb composition. Such behavior is expected as the mixing between the EB, HH, and LH bands enhances the $g$-factor, and the mixing is strongly correlated with the size of the band gap. Therefore, the $g$-factors and band gaps exhibit opposite bowing effects. However, the bowing in $g_{\textrm{exp}}$ is found smaller than  that in $g_{\textrm{theory}}$. As discussed before, this is because the Roth formula is a single-band theory and fails to handle the orbital mixing effect as band gap reduces \cite{dresselhaus2007group}. 

\begin{figure}[t]
\includegraphics[width=0.48\textwidth] {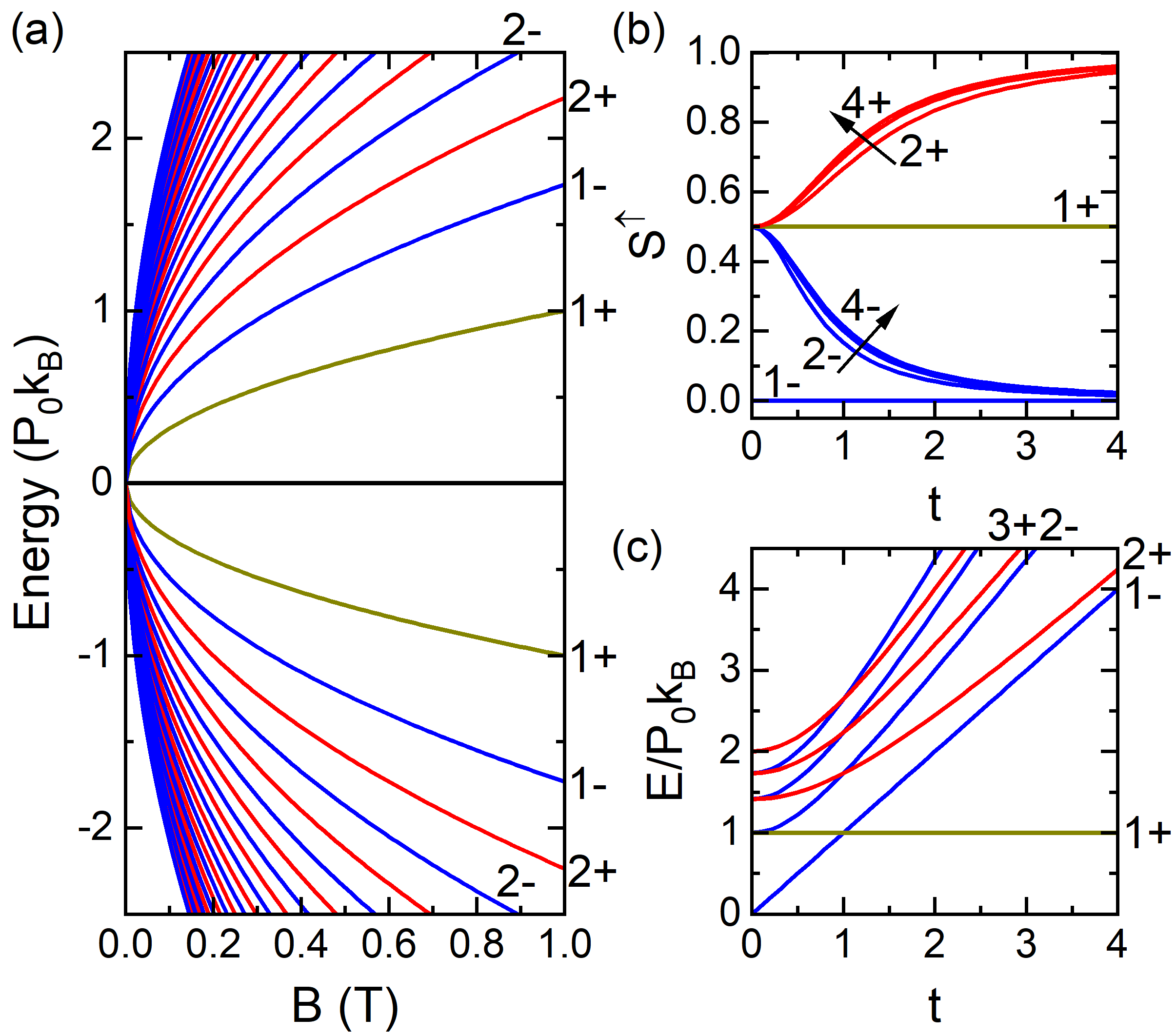}
\caption{(a) Landau fan diagram of a triply degenerated band structure (i.e., zero band gap) with $t=\sqrt{3}$ for the case of zinc-blende semiconductors. The energy is in units of $P_0 k_0$, where $k_0=\sqrt{e/\hbar}$. (b) The spin up component in low-lying LLs as a function of $t$. (c) The $t$ dependence of the low-lying LL energies. In all panels, the red and blue colors denote the spin up and spin down component dominant LLs, respectively. The dark yellow color denotes the LLs with equally mixed opposite spins. The black line denotes the highly degenerated HH LLs.}
\label{Triple}
\end{figure}

Further enhancement of the $g$-factor is possible when the band gap approaches zero. In this case, the EB, HH, and LH bands are degenerated (forming a triple point), and their interactions become the dominant effect. For simplicity, as the SO band is still far from these bands, we can omit the SO band  presence in the following discussion. We thus arrive at the following Hamiltonian $H_{\pm}$
\begin{align*}
H_{+}=\begin{bmatrix}
0 & itU^{\dagger} & iU \\
-itU &0 & 0 \\
-iU^{\dagger} & 0 &0\\
\end{bmatrix},
H_{-}=\begin{bmatrix}
0 & -tU & -U^{\dagger} \\
-tU^{\dagger} &0 & 0 \\
-U & 0 &0\\
\end{bmatrix}.
\end{align*}
Here, $U=P_{0}k_{-}$, and for a more general discussion, we use $t$ to denote the ratio of the coupling strength between the EB and HH to that between the EB and LH. The corresponding LL energy reads
\begin{align*}
&E^{0}_{n,\pm}=0, \quad n=0,2,3,4... \\
&E_{n,+}^{\alpha}=\alpha P_0k_B \sqrt{n(1+t^2)-t^2}, \quad n=1,2,3,4...\\ &E_{n,-}^{\alpha}=\alpha P_0k_B \sqrt{n(1+t^2)-1}, \quad n=1,2,3,4....
\end{align*}
where $k_B=\sqrt{eB/\hbar}$, and $e$ is the elementary charge. Each LL has three indices. The superscript $\alpha$ is the band index and takes the value of $0,+1,-1$, denoting the HH, EB, and LH bands, respectively. The first subscript $n$ denotes the LL index in each band, and the second subscript $\pm$ denotes the subblock $H_{\pm}$ to which the eigenstate relates. Figure \ref{Triple}(a) shows the magnetic field dependence of the calculated LL energies with $t=\sqrt{3}$ , which is the case for III-V semiconductors. Due to the electron-hole symmetry (i.e., $E^{-1}_{n,\pm}=-E^{+1}_{n,\pm}$), we will focus on the $\alpha=+1$ LLs in the discussion below. We will also exclude the discussion of the spin states in the $\alpha=0$ LLs as their Zeeman effect is negligible due to large degeneracy. In this case, we can omit the band index for simplicity.

%%%%% FIG. 4 %%%%%
%%%%%%%%%%%%%%

As the basis state for each subblock $H_{\pm}$ is not a pure spin state, the spin up component of a LL is found to be
\begin{align*}
S_{n,+}^{\uparrow}=1-\frac{n/2}{n(1+t^2)-t^2}, \quad
S_{n,-}^{\uparrow}=\frac{(n-1)/2}{n(1+t^2)-1}.
\end{align*}
Figure \ref{Triple}(b) shows the calculated spin up component of the low-lying LLs as a function of $t$. We find that independent of $t$, LL$_{1,+}$ is equally spin mixed while LL$_{1,-}$ is fully spin down polarized. For other LLs, they become more spin polarized with increasing $t$. Hence, for $t$ that gives decent spin polarization, the Zeeman splitting is now directly connected to the orbital energy levels (i.e., the LLs) and exhibits a peculiar relativistic $\sqrt{B}$ magnetic field dependence (Fig. \ref{Triple}(a)), in stark contrast to the conventional linear in $B$ Zeeman splitting. 

On the other hand, the magnitude of the Zeeman splitting also depends on the choice of $t$. Figure \ref{Triple}(c) shows the $t$ dependence of the low-lying LL energies. For $t=0$, 1, and $t \rightarrow +\infty$, the LLs of opposite dominant spin components are degenerated, and thus zero Zeeman effect. On the contrary, when a LL is equally separated from two neighboring LLs of opposite spins, the optimized Zeeman effect is achieved. For example, a simple calculation using the relation $E_{2,-}-E_{2,+}=E_{2,+}-E_{1,-}$ gives an optimized $t \approx 1.7$ for large Zeeman splitting in LL$_{2,+}$, which is close to $t=\sqrt{3}$ in III-V semiconductors. The optimized $t$ for other LLs is also close to this value. 

It is interesting to compare the Zeeman effect in such triple point semimetals to those of Dirac semimetals such as graphene \cite{Jiang2019valley} and ZrTe$_5$ \cite{wang2021magneto}. In the two-band model (as in Dirac semimetals), the interaction between the two bands leads to degenerated LLs with no dominant spin components. This is equivalent to taking $t \rightarrow 0$ or $+\infty$ in our model, where no Zeeman effect exists if only considering the orbital interaction. The Zeeman effect comes into play through the interaction with remote bands \cite{jiang2022giant,song2019principle,dresselhaus2007group}, which leads to a relatively small $g$-factor. However, in triple point semimetals, the additional interaction with the third band can lift the degeneracy of the LLs (except for the lowest two LLs). Therefore, the Zeeman effect can reveal itself through the splitting of the orbital energy levels and no longer take effect through perturbations. In this case, the $g$-factor can be more easily and effectively manipulated through the interactions between the three bands (EB, HH, and LH) rather than with the remote bands. These observations could be useful in designing high $g$-factor in future topological materials.

Before closing, we comment on how to enhance the Zeeman effect in practicable materials. We find that $t=\sqrt{3}$ is an ideal ratio which gives rise to a decent 80\% spin polarization in $n>1$ LLs as well as the ideal energy spacing between spin polarized LLs. In fact, this ratio is protected by the crystal symmetry and hence it can be also applied to the zinc-blende type semiconductor \cite{voon2009kp}. Using a typical value of $E_P=20$ eV, the Zeeman splitting for LL$_{2,+}$ is about 11 meV at 1 T (i.e., $\min \{ E_{2,-}-E_{2,+},E_{2,+}-E_{1,-}\}\approx 11$ meV), which corresponds to an effective $g$-factor of $g \approx -200$. Our finding is consistent with that reported on triple point (zinc-blende) HgCdTe \cite{orlita2014observation}. Therefore, zinc-blende type semiconductors with zero energy gap are ideal candidates for realizing large Zeeman effects.

This work was primarily supported by the NSF (grant nos. DMR-1809120 and DMR-1809708). The MBE growth at Stony Brook was also supported by the U.S. Army Research Office (Grant No. W911NF2010109) and the Center of Semiconductor Materials and Device Modeling. The magneto-IR measurements were performed at the National High Magnetic Field Laboratory, which is supported by the NSF Cooperative Agreement (nos. DMR-1644779 and DMR-2128556) and the State of Florida. S.M., D.S., and Z.J. acknowledge support from the DOE (for magneto-IR) under grant no. DE-FG02-07ER46451. Y.J. acknowledges support from the National Natural Science Foundation of China (Grant No. 12274001) and the Natural Science Foundation of Anhui Province (Grant No. 2208085MA09).

%\bibliographystyle{apsrev4-1}
%\bibliography{spin_ref}


\end{document}